\documentstyle[equations,12pt]{article}
\renewcommand{\theequation}{\thesection .\arabic{equation}}
\newtheorem{Corollary}{Corollary}[section]

\begin{document}

\title{Fractional Darboux Transformations}

\author{Mayer Humi\\
Department of Mathematical Sciences\\
Worcester Polytechnic Institute\\
100 Institute Road\\
Worcester, MA  01609}

\maketitle
\thispagestyle{empty}

\begin{abstract}
In this paper we utilize the covariance of Riccati equation with
respect to linear fractional transformations to define classes of conformally
equivalent second order differential equations. This motivates then
the introduction of fractional Darboux transformations which can be
recognized also as generalized Cole-Hopf transformations.  We apply
these transformations to find Schrodinger equations with isospectral
potentials and to the linearization of some new classes of nonlinear
partial differential equations.
\end{abstract}

\newpage 

\setcounter{page}{1}
\section{Introduction}

Darboux transformations [1,2,3] of the form
\begin{equation}
\label{1.1}
\psi = \left[A(x) + B(x) \frac{\partial}{\partial x}\right]\phi(x)
\end{equation}
have many different applications in mathematical physics [9,10,11,12,13]
and the theory of differential
equations.  Among these one can find the Factorization method (raising
and lowering operators for one or several coupled differential
equations [1,4]) separation of coupled Schrodinger equations [6] and
applications to nonlinear integrable equations.  Furthermore with
proper generalization this transformation has been applied to
multidimensional problems [6,10,14,16] and discrete systems [15,17].

In its simplest context the transformation (\ref{1.1}) ``connects'' the
solutions of two Schrodinger equations with different potentials
$u(x), v(x)$ i.e.
\begin{equation}
\label{1.2}
\phi^{\prime\prime} = (u(x) + \lambda)\phi
\end{equation}
\begin{equation}
\label{1.3}
\psi^{\prime\prime} = (v(x) + \lambda)\psi.
\end{equation}
Letting $B(x) = 1$  one can easily show that in order for eqs
(\ref{1.2}), (\ref{1.3}) to be related by the transformation
(\ref{1.1}) $A(x), u(x), v(x)$ must satisfy;
\begin{equation}
\label{1.4}
A^{\prime\prime} + u^\prime + A(u-v) = 0
\end{equation}
\begin{equation}
\label{1.5}
2A^\prime + u - v = 0
\end{equation}
Eliminating  $(u-v)$ between these equations and integration yields
\begin{equation}
\label{1.6}
A^\prime - A^2 + u = -\nu
\end{equation}
where  $\nu$ is an integration constant.  Eq. (\ref{1.6}) is a
Riccati equation which can be linearized by the transformation  $A =
-\zeta^\prime/\zeta$  which leads to
\begin{equation}
\label{1.7}
\zeta^{\prime\prime} = (u(x) + \nu)\zeta.
\end{equation}
Thus $\zeta$ is an eigenfunction of the original eq. (\ref{1.2}) with
$\lambda = \nu$.  From (\ref{1.5}) we now infer that
\begin{equation}
\label{1.8}
v = u - 2(ln\zeta)^{\prime\prime}
\end{equation}
i.e. a Darboux transformation changes the potential function  $u(x)$
by  $\Delta u = -2(ln\zeta)^{\prime\prime}$  where  $\zeta$ is an
arbitrary eigenfunction of (\ref{1.2}).

Our objective in this paper is to introduce fractional Darboux
transformations which are defined as transformations of the form
\begin{equation}
\label{1.9}
\psi = \displaystyle\frac{A(x)\phi(x) + B(x)\phi^\prime}{C(x)\phi(x) +
D(x)\phi^\prime}
\end{equation}
and elaborate on some of their applications to linear and nonlinear
differential equations.  To {\it motivate} the introduction of these
transformations we use the covariance of the Riccati equation with
respect to linear fractional transformations.  We then show that this covariance
induces an equivalence relation between different second order
differential equations whose solutions are related by a transformation
of the form (\ref{1.9}) (Sec. 2).  In Section 3 we consider the
classification problem for these ``conformally
equivalent'' equations.  In Section 4 and 5 we explore then the application of
the transformation (\ref{1.2}) to Schrodinger equations and nonlinear
equations.  We end with some observations and conclusions in
Section 6.

\setcounter{equation}{0}
\section{Conformally Equivalent Equations}

Consider a linear second order differential equation
\begin{equation}
\label{2.1}
p(x)w^{\prime\prime} + q(x)w^\prime + r(x)w = 0
\end{equation}
(with smooth enough coefficients).

Introducing
\begin{equation}
\label{2.2}
y = \displaystyle\frac{w^\prime}{w}
\end{equation}
eq. (\ref{2.1}) is transformed into a Riccati equation
\begin{equation}
\label{2.3}
y^\prime = -y^2 - \displaystyle\frac{q(x)}{p(x)}\;y -
\displaystyle\frac{r(x)}{p(x)}.
\end{equation}
However it is well known that Riccati equation is ``covariant'' with
respect to the linear fractional transformation [5,7]
\begin{equation}
\label{2.4}
y = \displaystyle\frac{\alpha(x)z(x) + \gamma(x)}{\beta(x)z(x) +
\delta(x)}
\end{equation}
i.e. such a transformation takes a Riccati equation into another one
in its class
\begin{equation}
\label{2.5}
z^\prime = F(x) z^2 + G(x)z + H(x)
\end{equation}
In particular for eq. (\ref{2.3}) we have;
\begin{equation}
\label{2.6}
F = -[\alpha^2 + \alpha
\beta\left(\displaystyle\frac{q(x)}{p(x)}\right) +
\beta^2\left(\displaystyle\frac{r(x)}{p(x)} \right) +
(\alpha^\prime\beta - \alpha \beta^\prime)]/\Delta
\end{equation}
\begin{equation}
\label{2.7}
G = -[2\alpha\gamma + (\alpha \delta + \beta
\gamma)\left(\displaystyle\frac{q(x)}{p(x)}\right) +
2\beta\delta\left(\displaystyle\frac{r(x)}{p(x)}\right)
+(\alpha^\prime\delta - \alpha \delta^\prime) + (\gamma^\prime\beta -
\gamma\beta^\prime)]/\Delta
\end{equation}
\begin{equation}
\label{2.8}
H = -[\gamma^2
+\gamma\delta\left(\displaystyle\frac{q(x)}{p(x)}\right) +
\delta^2\left(\displaystyle\frac{r(x)}{p(x)}\right) +
(\gamma^\prime\delta - \gamma \delta^\prime)]/\Delta
\end{equation}
where  $\Delta = \alpha\delta - \beta \gamma \neq 0$.

However we can transform eq. (\ref{2.5}) back to a second order linear
differential equation by the transformation
\begin{equation}
\label{2.9}
z = \displaystyle\frac{-u^\prime(x)}{F(x)u(x)}
\end{equation}
to obtain
\begin{equation}
\label{2.10}
u^{\prime\prime} - \left(G(x) +
\displaystyle\frac{F^\prime(x)}{F(x)}\right)u^\prime + H(x)F(x)u = 0.
\end{equation}
We conclude then that the sequence of transformations (\ref{2.2}),
(\ref{2.4}), (\ref{2.9}) establishes an equivalence relation between
linear second order differential equations and we shall say that such
equations are ``conformally equivalent''.

To see how the solutions of eqs (\ref{2.2}), (\ref{2.10}) are related
we note that from (\ref{2.2}), (\ref{2.4}) we have
\begin{equation}
\label{2.11}
z = \displaystyle\frac{\delta w^\prime - \gamma w}{\alpha w - \beta
w^\prime}
\end{equation}
and hence using (\ref{2.9})
\begin{equation}
\label{2.12}
u(x) = exp\left(\int F(x)\displaystyle\frac{\gamma w - \delta
w^\prime}{\alpha w - \beta w^\prime} dx \right).
\end{equation}
We conclude then that up to exponentiation and integration the
relationship between the solutions is given by a Fractional Darboux
transformation.
\begin{equation}
\label{2.13}
\tilde{w}(x) = \displaystyle\frac{A(x)w + B(x)w^\prime}{C(x)w + D(x)w^\prime}
\end{equation}
We now explore some properties of this relationship when
$\alpha,\beta,\gamma,\delta$ is eq. (\ref{2.4}) are constants.  Under
this restriction it is well known that the transformation (\ref{2.4})
is equivalent to a sequence of three transformations
\begin{subequations}
\label{214}
\begin{eqnarray}
y &=& a_1z_1 + b_1  \label{214a}\\
z_1 &=& \frac{1}{z_2} \label{214b}\\
z_2 &=& a_2z_3 + b_2 \label{214c}
\end{eqnarray}
\end{subequations}
Since affine transformations (\ref{214a}), (\ref{214c}) induce only a
minor change in the original equation (they just add constant to
$r(x)$) it is interesting to find those equations which remain
invariant with respect to the transformation  $y = \frac{1}{z}$.  To
this end we let  $p(x), q(x), r(x)$ be polynomials in  $x$
\begin{equation}
\label{2.15}
p(x) = \displaystyle\sum^n_{k=0} p_kx^k,\;\;q(x) =
\displaystyle\sum^n_{k=0} q_kx^k,\;\;r(x) = \displaystyle\sum^n_{k=0}
r_kx^k
\end{equation}
and solve the equations
\begin{equation}
\label{2.16}
\displaystyle\frac{q(x)}{p(x)} = -\left(G(x) +
\displaystyle\frac{F^\prime(x))}{F(x)}\right)
\end{equation}
\begin{equation}
\label{2.17}
\displaystyle\frac{r(x)}{p(x)} = H(x)F(x)
\end{equation}
for the coefficients  $p_k,q_k,h_k$.  When  $n=2$ we find that up to
translations in  $x$ (i.e. $\bar{x} = x + a)$ there are only three
equations with this property
\begin{itemize}
\item[(1)] 
\begin{equation}
\label{2.18} w^{\prime\prime} - n^2w = 0
\end{equation}
\item[(2)] Chebychev equation
\begin{equation}
\label{2.19}
(p_0 - x^2)w^{\prime\prime} - xw^\prime + h_0w = 0
\end{equation}

\item[(3)] \begin{equation}\label{2.20} x\left(x-2\;\frac{q_0}{p_0}\right)\;
w^{\prime\prime} + \frac{q_0}{p_0}\; w^\prime + \frac{h_0}{p_0}\; x^2 w =
0
\end{equation}
\end{itemize}
Letting  $p_0 = 1,\;\;q_0 = \frac{1}{2}$ in (\ref{2.20}) we have
$$
x(1 - x)w^{\prime\prime} - \frac{1}{2} w^\prime - h_0 x^2 w = 0
$$
This equation does not appear in Kamke's book [5].  However it belongs to
the class of Hill-Mathieu equations through the transformation  $x =
\cos\theta$.

In Table 1 we present the effect of the transformation  $y =
\frac{1}{z}$ on several classes of equations whose solutions are the
Special Functions of Mathematical Physics.  We would like to emphasize
however that in general  $\alpha,\beta,\gamma,\delta$ in eq. (\ref{2.4})
can be functions of  $x$  rather than just constants and consequently
the conformal equivalence class of each of these equations can be very
large indeed.  

Finally for the Schrodinger equation (\ref{1.2}) we find that when
$\alpha,\beta,\gamma,\delta$ are constant with  $\alpha\delta-\beta
\gamma = 1,\;\;Q(x),\;\;R(x)$ in eq. (\ref{2.10}) take the following form;
\begin{equation}
\label{2.21}
Q(x) = 2\alpha\gamma - 2\beta\delta(u(x) + \lambda) -
\displaystyle\frac{\beta^2u^\prime(x)}{\beta^2(u+\lambda)-\alpha^2}
\end{equation}
\begin{equation}
\label{2.22}
R(x) = [\delta^2(u+\lambda)-\gamma^2][\beta^2(u+\lambda)-\alpha^2]
\end{equation}

\setcounter{equation}{0}
\section{The Classification Problem}

To use the algorithm described in the previous section in a practical
manner one must be able to determine when a differential equation
\begin{equation}
\label{omega}
w^{\prime\prime}_1 + q_1(x)w^\prime_1 + r_1(x)w_1 = 0
\end{equation}
is conformally equivalent to eq. (\ref{2.1}).  To this end one must
solve the equations
\begin{equation}
\label{frac}
q_1(x) = G(x) + \frac{F^\prime(x)}{F(x)}
\end{equation}
\begin{equation}
\label{gamma}
r_1(x) = H(x)F(x)
\end{equation}
for appropriate  $\alpha(x), \beta(x), \gamma(x), \delta(x)$ subject
to the condition  $\Delta \neq 0$.

To make progress towards the solution of this classification problem
we shall consider two separate cases;  $\beta(x) \neq 0$ and
$\beta(x) = 0$.

\vspace*{.05in}

\noindent A.\,\,\,$\beta(x) \neq 0$.

In this case we can assume without loss of generality that  $\beta(x)
= 1$.  The resulting expression for  $q_1(x)$ can be used to solve for
$\gamma(x)$ in terms of  $q_1(x), \alpha(x), \delta(x)$ and the
coefficients of eq. (2.1).  Substituting this expression for  $\gamma(x)$ in
eq. (\ref{3.3})(with  $\beta(x) = 1)$ we obtain;
\begin{equation}
\label{beta}
r_1(x) = \frac{q_1(x)^2}{4} + \frac{1}{2} \frac{dq_1(x)}{dx} + R_1(x)
\end{equation}
where  $R_1(x)$ which depends only on  $\alpha(x)$ and the coefficients
of eq. (\ref{2.1}) is given by eq. (A.1).  (To simplify this equation
we let  $p(x) = 1)$.

Eq. (\ref{beta}) is a single ODE for  $\alpha(x)$ and if a solution to
this equation exists then equations (\ref{2.1}), (\ref{omega}) are
conformally equivalent.  However in general eq. (\ref{beta}) is
nonlinear (and intractable) equation for $\alpha(x)$.  Due to this
circumstance it is appropriate to use eq. (\ref{beta}) in a
``reverse'' manner viz. assume a fixed functional form for
$\alpha(x)$ and determine the corresponding form of  $R_1(x)$.  This expression can be evaluated for
various classes of 2nd order ODEs and an appropriate table can provide
a quick reference to determine if a given equation is conformally
equivalent to any of the classes that appear in the table (under the
restrictions imposed on  $\alpha(x))$.

Table 2 presents the different  $R_1(x)$ that correspond to some classes
of special functions when  $\alpha(x) = 0$. 

From this table we gather the following:
\begin{Corollary}
\begin{enumerate}
\item $\omega^{\prime\prime} + \omega = 0$
is conformally equivalent to Bessel equation of order $1/2$
\item Bessel equations of order  $0,1$ are conformally equivalent to
each other.
\end{enumerate}
\end{Corollary}

\noindent B.\,\,\,$\beta(x) = 0$

As can be expected this is a simple case to treat.

To begin with we can assume without loss of generality that
$\delta(x) = 1(\alpha(x) \neq 0)$ and obtain for  $q_1(x), r_1(x)$ the
following expressions (we let  $p(x) = 1$)
\begin{equation}
\label{let}
q_1(x) = 2\alpha(x) + q(x)
\end{equation}
\begin{equation}
\label{and}
r_1(x) = \alpha^2(x) + \alpha(x)q(x) + r(x) + \frac{\partial
\alpha(x)}{\partial x}.
\end{equation}
Solving (\ref{let}) for  $\alpha(x)$ and substituting in (\ref{and})
we obtain eq. (\ref{beta}) with  $R_1(x)$ given by
\begin{equation}
\label{given}
R_1(x) = -\frac{q(x)^2}{4} - \frac{1}{2}\;\frac{dq(x)}{dx} + r(x)
\end{equation}
(observe that  $R(x)$ is independent of  $\alpha(x))$.

To recast this algorithm in perspective we note that Forbenius method
expresses the solution of (\ref{2.1}) in terms of powers of $x$.  The
algorithms described above enable us to approach this problem
differently by asking whether the solution of (\ref{omega}) can be
expressed naturally in terms of other classes of function
(e.g. special functions) i.e. when (\ref{omega}) is conformally
equivalent to an equation whose solution is already known.

\setcounter{equation}{0}
\section{Application to Schrodinger Equation}

In this section we consider the application of fractional Darboux
transformations in the form
\begin{equation}
\label{3.1}
\psi(x) = \displaystyle\frac{A(x)\phi(x) + \phi^\prime(x)}{B(x)\phi(x) +
\phi^\prime(x)}
\end{equation}
to relate the solutions of equations (\ref{1.2}), (\ref{1.3}).
Differentiating (\ref{3.1}) twice and substituting in (\ref{1.3})
using (\ref{1.2}) to eliminate the higher order derivatives of
$\phi(x)$ we obtain a polynomial which contains the monomials $\phi^3,
\phi^2 \phi^\prime, \phi(\phi^\prime)^2$ and $(\phi^\prime)^3$.  To annul this
polynomial we let the coefficient of each of these monomials be zero.
By using the last three equations to express  $\lambda, v(x)$ and
$\displaystyle\frac{du}{dx}$ and simplifying we obtain from the first
equation (coefficient of  $\phi^3$)
\begin{equation}
\label{3.2}
[(B-A)^\prime - (B^2-A^2)]^\prime - 2B[(B-A)^\prime - (B^2-A^2)] = 0.
\end{equation}
This can be solved by the Ansatz
\begin{equation}
\label{3.3}
(B-A)^\prime - (B^2-A^2) = 0
\end{equation}
substituting this relation in the equations for $\lambda, v$ and
$\displaystyle\frac{du}{dx}$
yield the following equations
\begin{equation}
\label{3.4}
u(x) = A^2 - A^\prime - c
\end{equation}
\begin{equation}
\label{3.5}
v(x) = 2A(A-B) - c
\end{equation}
\begin{equation}
\label{3.6}
\lambda = c
\end{equation}
where  $c$ is a constant of integration.  Thus
\begin{equation}
\label{3.7}
\Delta u = v(x) - u(x) = A(A-2B) + A^\prime.
\end{equation}
To use these equations in practical applications we start with
eq. (4.4) to find $A$.  This Riccati equation can be converted
to
\begin{equation}
\label{3.8}
\zeta^{\prime\prime}_1 = (u(x) + c)\zeta_1
\end{equation}
by the transformation
\begin{equation}
\label{3.9}
A = -\displaystyle\frac{\zeta^\prime_1}{\zeta_1}.
\end{equation}
It follows then that  $\zeta_1$ is a solution of eq. (\ref{1.2}) with
$\lambda = c$.  Similarly if we substitute for  $A$ in (\ref{3.3})
using (\ref{3.4}) we obtain
\begin{equation}
\label{3.10}
B^\prime - B^2 = -u - c.
\end{equation}
Thus  $\zeta_2$ $\left( \mbox{where}\;\;  B =
-\displaystyle\frac{\zeta^\prime_2}{\zeta_2}\right)$ is also a solution of
(\ref{1.2}) with the same eigenvalue.  It follows then that
\begin{equation}
\label{3.11}
\Delta u = [(ln\zeta_1)^\prime]^2 - 2(ln\zeta_1)^\prime(ln\zeta_2)^\prime -
(ln\zeta_1)^{\prime\prime}
\end{equation}
which is completely different from (\ref{1.8}) for Darboux
transformation (\ref{1.1}).

\setcounter{equation}{0}
\section{Generalization of Cole-Hopf Transformation}

It is well known that Cole-Hopf transformation [8]
\begin{equation}
\label{4.1}
\psi(x,t) = -2\nu\displaystyle\frac{\phi_x(x,t)}{\phi(x,t)}
\end{equation}
can be used to linearize the Burger's equation
\begin{equation}
\label{4.2}
\psi_t + \psi\psi_x = \nu\psi_{xx}
\end{equation}
and transform it to the heat equation
\begin{equation}
\label{4.3}
\phi_t = \nu\phi_{xx}.
\end{equation}
It is now easy to recognize that the fractional Darboux transformation
\begin{equation}
\label{4.4}
\psi(x,t) = \displaystyle\frac{A(x)\phi(x,t) + C(x)\phi_x(x,t)}{B(x)\phi(x,t) +
D(x)\phi_x(x,t)} = \frac{N}{E}
\end{equation}
is actually a generalization of (\ref{4.1}).  It is natural therefore
to apply it to the linearization of nonlinear partial differential
equations.

In the following we restrict ourselves to the classification of
nonlinear equations which can be transformed to the heat equation
through the transformation (\ref{4.4}).

To this end we compute
$$
\displaystyle\frac{\partial \psi}{\partial t} -
\frac{\partial^2\psi}{\partial x^2}
$$
using (\ref{4.4}) and simplify the expressions thus obtained by
replacing  $\phi_t$ by $\phi_{xx}$ at each step (we set  $\nu = 1$ in
eq. (\ref{4.3})).

After some algebra we obtain
\begin{eqnarray}
\label{4.5}
\psi_t - \psi_{xx} &=& \displaystyle\frac{\left[A^{\prime\prime}+(2A^\prime
+
C^{\prime\prime})\left(\frac{\phi_x}{\phi}\right)+C^\prime\left(\frac{\phi_{xx}}{\phi}\right)\right]}{B+D\left(\frac{\phi_x}{\phi}\right)}\\ \nonumber
&+&
\psi\displaystyle\frac{\left[B^{\prime\prime}+(2B^\prime+D^{\prime\prime})\left(\frac{\phi_x}{\phi}\right)+2D^\prime\left(\frac{\phi_{xx}}{\phi}\right)\right]}{B
+ D\left(\frac{\phi_x}{\phi}\right)} 
+ 2\psi_x\frac{E_x}{E}.
\end{eqnarray}
However from (\ref{4.4})
\begin{equation}
\label{4.6}
\displaystyle\frac{\phi_x}{\phi} = \frac{B\psi - A}{C - D\psi}.
\end{equation}
Substituting(\ref{4.6}) and using the identity
\begin{equation}
\label{4.7}
\displaystyle\frac{\phi_{xx}}{\phi} =
\left(\frac{\phi_x}{\phi}\right)_x +
\left(\frac{\phi_x}{\phi}\right)^2.
\end{equation}
in  eq. (\ref{4.5}) we obtain for the right hand side an expression
which contains monomials of  $\psi, \psi_x$ up to the fourth and
second degree respectively.  However if we restrict ourselves to
second order nonlinearity it is enough to let  $D = 0$ and set $B(x) =1$.  The simplified form of eq. (\ref{4.5}) in
this case is
\begin{eqnarray}
\label{4.8}
C^2(\psi_t-\psi_{xx}) &=& -C^\prime\psi^2 + [C^\prime(2A+C^\prime) -
C(C^\prime-2A)^\prime + 2C\psi_x]\psi \nonumber \\
&-& C(C^\prime + 2A)\psi_x + AC(C^\prime + 2A)^\prime -
A(C^{\prime\prime} + C^\prime A)\nonumber \\
&+& C(C^\prime A^\prime - A^{\prime\prime}C).
\end{eqnarray}
In particular if  $A,B,C,D$ are constants
eq. (\ref{4.5}) simplifies and we obtain
\begin{equation}
\label{4.9}
\psi_t - \psi_{xx} =
2\psi_x\displaystyle\frac{A-B\psi-D\psi_x}{D\psi-C}
\end{equation}
which is a generalization of Burger's equation.

Similarly we can use the transformation (5.1) (with $A,B,C,D$ being
functions of  $x$ only) to relate a linear
second order equation of the form (\ref{2.1}) (with $w$ replaced by $\phi$) to second order
nonlinear ODEs.  If we restrict ourselves to nonlinearities up to the
third order we find that we must set  $D(x) = 0$ (and without loss
generality let $B(x) = 1)$.  The general form of the resulting
nonlinear second order equation is

\begin{eqnarray}
C^2\psi^{\prime\prime} &=& 2\psi^3-(3\,q\; C + 6\,A + 2\,C^\prime)\psi^2\nonumber \\
&+&[(-2\,r + q^\prime +q^2)C^2 +
(C^{\prime\prime}+2\,q\; C^\prime +6\,A
q) C \nonumber \\
&+& 6\,A^2+4\,C^\prime A] \psi+(r^\prime
+q\;r)C^3\\
&+&[(-q^\prime-q^{2}+2\,r)A + A^{\prime\prime}+2\,C^\prime r]C^2 \nonumber \\
&+&[-3\,q A^2 - (C^{\prime\prime} + 2\,q C^\prime)A]C
-2\,A^3-2\,C^\prime A^2\nonumber
\end{eqnarray}

 Similar algorithm can be used to find other nonlinear equations that
can be transformed to a given linear equation via a transformation of
the form (\ref{4.4}).

\section{Conclusions}

We introduced in this paper Fractional Darboux transformations and
elaborated on some of their applications.  In view of the wide range
of applications that regular Darboux transformations have in the
literature we expect that this new transformation will generalize many
known results.  In particular the systematic adaption of this new
transformation to the computation of Laddar operators for new
potentials in the spirit of [1,4] has yet to be worked out.  We hope to elaborate on these subjects
in subsequent publications.
\newpage

\setcounter{equation}{0}
\def\theequation{A. \arabic{equation}}
\begin{center}
{\bf Appendix}
\end{center}

In this appendix we give the general form of  $R_1(x)$ in
eq. (3.4).  
\begin{equation}
R_1(x) = \displaystyle\frac{N_1(x)}{D_1(x)}
\end{equation}
where

\begin{eqnarray}
N_1 &=& 2(\alpha^2+\alpha q + \alpha^\prime
+r)\alpha^{\prime\prime\prime} -
3(\alpha^{\prime\prime})^2-6(2\alpha+(q\alpha)^\prime
+r^\prime)\alpha^{\prime\prime}\nonumber \\
&+&
12(\alpha^\prime)^3+6(4r+q^\prime-q^2)(\alpha^\prime)^2+[4(4r-2q^\prime-q^2)\alpha^2
\nonumber \\
&+& 2(q^{\prime\prime}+8qr-8r^\prime-2q^3)\alpha
+8r(2r+q^\prime)-4q(qr+2r^\prime)+2r^{\prime\prime}]\alpha^\prime
\nonumber \\
&+&
(4r-2q^\prime-q^2)\alpha^4+2(q^{\prime\prime}-2r^\prime-q^3+4qr-q^\prime
q)\alpha^3\\
&+& [8r^2-q^4+2q^2r
+(2q^{\prime\prime}-6r^\prime)q-3q^{\prime\prime}+2r^{\prime\prime}]\alpha^2+[8qr^2
\nonumber \\
&+&2(q^{\prime\prime}+q^\prime
q-2r^\prime-q^3)r+2r^{\prime\prime}q-2q^2r^\prime-6r^\prime
q^\prime]\alpha \nonumber \\
&+&
4r^3+(2q^\prime-q^2)r^2+(2r^{\prime\prime}-2qr^\prime)r-3(r^\prime)^2\nonumber
\end{eqnarray}
\begin{equation}
D_1 = 4[\alpha(\alpha +q)+\alpha^\prime + r]
\end{equation}

For brevity we suppressed the dependence on  $x$ in
eqs. (A.2), (A.3).

\newpage
\centerline{\bf References}

\begin{enumerate}
\item J.G. Darboux - C. R. Acad. Sci. Paris {\bf 94} p. 1456 (1882)
\item L. Infeld and T.E. Hull - Rev. Mod. Phys. {\bf 23} p. 2 (1951)
\item M.M. Crum - Q. J. Math. {\bf 2} p. 121 (1955)
\item M. Humi - J. Math. Phys. {\bf 26} p. 26 (1986)
\item E. Kamke - Differential Gliechungen, Chelsea Pub. Co., NY (1948)
\item V.M. Strelchenya - J. Phys. A {\bf 24} p. 4965 (1991)
\item M. Humi - J. Phys. A {\bf 18} p. 1085 (1985)
\item E. Hopf - Comm. Pure Appl. Math. {\bf 3} p. 201 (1950)
\item W.M. Zhang - J.Math. Phys. {\bf 25} p. 88 (1984)
\item P.C. Sabatier - Inverse Problems {\bf 14} p. 355 (1998)
\item P.G. Estevez - J. Math. Phys. {\bf 40} p. 1406 (1999)
\item B. Bagchi and A. Ganguly - Inter. J. Mod. Phys. {\bf 13} p. 3711
(1998)
\item N.V. Ustinov - Rep. Math. Phys. {\bf 46} p. 279 (2000)
\item A. Gonzalez-Lopez and N. Kamran - J. of Geometry \& Physics,
{\bf 26} p. 202 (1998)
\item J.J.C. Nimmo - Chaos, Solitons \& Fractals {\bf 11} p. 115
  (2000)
\item R. Klippert and H.C. Rosu - Nuovo Cimento B {\bf 115} p. 350 (2000)
\item M.A. Reyes and H.C. Rosu - Nuovo Cimento B {\bf 114} p. 717-722 (1999)
\end{enumerate}

\newpage

\begin{table}[ht]

\caption{The effect of the sequence of transformations (2.2), (2.4),
(2.9) on the equations in the second column when (2.4) is given by  $y
= \frac{1}{z}$}

\vspace*{.15in}

\begin{tabular}{ccccccc}
Class of equations    &&&Original equation         && &Transformed
equation\\ \hline
\\
\vspace*{.15in}

Constant Coefficient  &&&$a\omega^{\prime\prime} +
b\omega^\prime+c\omega = 0$   &&&Same\\
\\
Legendre &&&$(1-x^2)w^{\prime\prime}-2xw^\prime +n(n+1)w=0$
&&&$(1-x^2)u^{\prime\prime} +n(n+1)u=0$\\
\\
Hermite  && &$w^{\prime\prime}-2xw^\prime +2nw=0$   &&&$u^{\prime\prime}
+2xu^\prime +2nu=0$\\
\\
Bessel  &&&$x^2w^{\prime\prime}+xw^\prime+(x^2-n^2)w=0$
&&&$u^{\prime\prime}-\frac{x^2+n^2}{x(x^2-n^2)}\;u^\prime +\frac{x^2-n^2}{x^2}u=0$\\
\\
Leguerre &&&$xw^{\prime\prime}+(1-x)w^\prime+nw=0$ &&&
$u^{\prime\prime}+u^\prime+\frac{n}{x} u=0$\\
\\
Chebyshev
&&&$(1-x^2)\omega^{\prime\prime}-x\omega^\prime + h^2\omega = 0$
&&&Same\\
\\
Hypergeometric &&&$x(1-x)w^{\prime\prime}+[\gamma-(\alpha+\beta
+1)x]w^\prime$
&&&$x(1-x)u^{\prime\prime}+[1-\gamma+(\alpha+\beta-1)x]u^\prime$\\
   \\
  &&&$-\alpha\beta w=0$  &&&$-\alpha\beta u=0$
\end{tabular}
\end{table}

\newpage

\begin{table}
\caption{Form of  $R_1(x)$ in eq. (3.4) for some classes of special
functions when  $\alpha(x) = 0$}

\begin{tabular}{c|c}\\
\\
    &$R_1(x)$ \\ \hline
\\
Constant coefficients    &$c - \frac{1}{4} b^2$\\
\\
Legendre	         &$\displaystyle\frac{n(n + 1)}{1 - x^2}$\\
\\
Hermite	                 &$-x^2 + 2n - 1$\\
\\
Bessel
    &$\displaystyle\frac{x^2[4x^4-3x^2(4n^2+1)+2n^2(6n^2-5)]+n^4(4n^2+1)}{4x^2(x^2-n^2)^2}$\\
\\
Leguerre         &$\displaystyle\frac{n}{x} - \displaystyle\frac{1}{4}$\\
\\
Chebyshev        &$-\displaystyle\frac{(4n^2-1)x^2-2(1+2n^2)}{4(x^2-1)^2n^4}$\\
\\
Hypergeometric
    &$\displaystyle\frac{[1-(\alpha-\beta)^2]x^2+2[(\alpha+\beta+1)\gamma-(2\alpha+1)\beta-(1+\alpha)]x+\gamma^2-1}{x^2(x-1)^2}$
\end{tabular}
\end{table}
\end{document}